\documentclass[journal,draftcls,onecolumn,12pt,twoside]{IEEEtranTCOM}
\usepackage{cite}
\usepackage{graphicx}
\usepackage{url}
\usepackage[centertags]{amsmath}
\newcommand{\beq}{\begin{equation}}
\newcommand{\eeq}{\end{equation}}
\newcommand{\beqn}{\begin{align}}
\newcommand{\eeqn}{\end{align}}

\begin{document}

\title{Improved Performance of RF Energy Powered Wireless Sensor Node with Cooperative Beam Selection}

\author{Tianqing~Wu,~
Hong-Chuan~Yang,~\IEEEmembership{Senior Member,~IEEE}
\thanks{This work was supported by a Discovery Grant from NSERC, Canada. T. Wu and H.-C. Yang are with the
Department of Electrical and Computer Engineering, University of Victoria, BC,
V8W 3P6, Canada (E-mail: $<$twu, hy@uvic.ca$>$).}}


\IEEEoverridecommandlockouts

\setcounter{page}{1}

\maketitle

\begin{abstract}
RF energy harvesting is a promising potential solution to provide convenient and perpetual energy supplies to low-power wireless sensor networks.
In this paper, we investigate the energy harvesting performance of
a wireless sensor node powered by harvesting RF energy from existing multiuser MIMO system.
Specifically, we propose a random unitary beamforming (RUB) based cooperative beam selection scheme to enhance the energy harvesting performance at the sensor. Under a constant total transmission power constraint, the multiuser MIMO system tries to select a maximal number of active beams for data transmission, while satisfying
the energy harvesting requirement at the sensor.
We derive the exact closed-form expression for the distribution function of harvested energy in a coherence time over Rayleigh fading channels.
We further
investigate the performance tradeoff of the average harvested energy at the sensor
versus
the sumrate of the multiuser MIMO system.
\end{abstract}


\section{Introduction}
\IEEEPARstart{W}{ireless} sensor networks (WSNs) have been used in a wide range of applications, such as environment monitoring, surveillance, health care, intelligent buildings and battle field control \cite{r1}.
The sensor nodes of WSN are usually powered by batteries with finite life time, which manifests as an important limiting factor to the functionality of WSN.
Replacing or charging the batteries may
either incur high costs for human labor or be impractical for certain application scenarios (e.g.applications that require sensors to be embedded into structures). 
Powering sensor nodes through ambient energy harvesting has therefore received a lot of attentions in both academia and industrial communities \cite{r2,r3}.
While various techniques have been developed to harvest energy from conventional ambient energy sources, such as solar power, wind power, thermoelectricity, and vibrational excitations \cite{n1,n2,n3,n4},
RF energy harvesting
has attracted a growing interest due to the intensive deployment of  cellular/WiFi wireless systems in addition to traditional radio/TV broadcasting systems \cite{r4}.
It has been experimentally proved that RF energy harvesting is feasible from the hardware implementation viewpoint \cite{hard1,hard2,hard3}. 



Previous literature on RF energy harvesting can be summarized as following.
The fundamental performance limits of simultaneous wireless information and energy transfer systems over point-to-point link were studied in \cite{unp1,unp2}.
A cognitive network that can harvest RF energy from the primary system is considered in \cite{mar}. The authors propose an optimal mode selection policy for sensor nodes to decide whether to transmit information or to harvest RF energy based on Markov modelling.
In \cite{opp}, the authors investigate mode
switching between information decoding and energy harvesting, based on the instantaneous
channel and interference condition over a point-to-point link.
A save-then-transmit protocol is proposed in \cite{ehref9} to minimize the outage probability of energy harvesting transmitters by finding the optimal time fraction for energy harvesting in a time slot, during which the wireless channel is assumed to be constant.
In most of these works, 
it is generally assumed that 
the channel gain remains constant during the whole energy harvesting circle, including obtaining channel state information, making decision accordingly, and then harvesting energy or decoding information.
In \cite{my paper}, the RF energy harvesting harvesting capability of wireless sensor node over multiple channel coherence time is characterized with consideration of hardware limitations, such as harvesting sensitivity and energy storage capacity, and interference from existing system.
It is shown in \cite{mimo} that
with channel information at the energy source node,
multi-antenna transmission can
can help increase the amount of harvested energy at the energy receiving node.
Inspired by \cite{mimo}, we consider a practical cooperative charging scenario
in this paper,
where
an existing multiuser MIMO system helps
the energy harvesting
of a
RF-energy-powered sensor node, while simultaneously serving its own users.

Random unitary beamforming (RUB) is a low-complexity transmission scheme for multiuser MIMO systems that requires very low feedback load, and has attracted continuing research interest \cite{lowcom, dpc}. 
It has been shown in \cite{lowcom} that if each user just feeds back its best beam index and the corresponding SINR, RUB can achieve the same sum-rate scaling law as the optimal dirty paper coding (DPC) transmission scheme \cite{dpc}.
However, most of previous works focus on data transmission for conventional RUB-based multiuser MIMO systems, and very limited work has considered RUB-based RF energy harvesting for coexisting networks.


With these observations in mind, we propose a RUB-based cooperative beam selection scheme, where the base station (BS) of the multiuser MIMO system selects a maximal number of active beams for transmission, while trying to satisfy energy harvesting requirement of the sensor, i.e., the harvested energy over each coherence time is above a predefined energy threshold.
With a constant total transmission power, the BS can enhance energy harvesting at the sensor by concentrating the transmission power on selected beams.
Meanwhile, the number of users that the BS can serve simultaneously is reduced.
To evaluate the performance tradeoff between the average harvested energy at the sensor and the sum-rate of the existing multiuser MIMO system,
we derive the closed-form statistical distribution of the amount of energy that can be harvested
with the proposed cooperative RF energy harvesting scheme.
These analytical results will help determine the optimal energy threshold value that can satisfy requirements of certain sensing applications, while considering the negative effect on the multiuser MIMO system.

The remainder of the paper is organized as follows. In Section \ref{smodel}, we briefly introduce the system and channel model, and mode of operation of our proposed RUB-based cooperative beam selection scheme under consideration. The performance analysis of the sensor node is investigated with numerical examples in Section \ref{scheme}. The performance analysis of the existing RUB-based multiuser MIMO system is investigated with numerical examples in Section \ref{secondary}.
Concluding remarks are given in Section \ref{conclude}.

\section{system and channel model}
\label{smodel}

\begin{figure}
\centering
\includegraphics[width=3in]{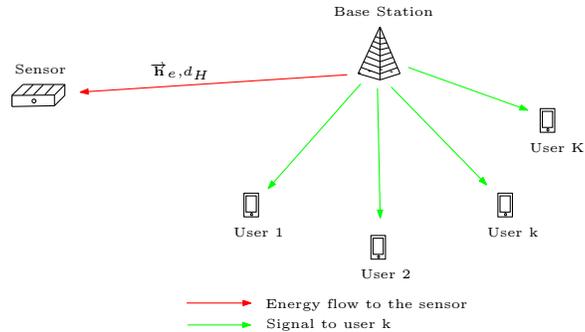}
\caption{System model for RUB-based cooperative RF energy harvesting.}
\label{model}
\end{figure}

\subsection{System Model}

We consider a single-antenna wireless sensor node
deployed in the coverage area of an existing RUB-based multiuser MIMO system, which could be cellular or WiFi systems.
The sensor
$\footnote{The sensor can also be a special user of multiuser MIMO system.}$
can harvest RF energy from the transmitted signal of the multiuser MIMO system, and use it as its sole energy source, as illustrated in Fig. \ref{model}. 
The multiuser MIMO system consists of single
BS
with $M$ antennas and $K$ $(K\ge M)$ single-antenna users.
The BS can serve up to $M$ selected users simultaneously using
random orthonormal beams generated from an isotropic distribution.
Let $\mathcal{W} = [\textbf{w}_1, \textbf{w}_2, \dots, \textbf{w}_M]^T$ 
denote the set of beam vectors, assumed to be known to both the BS and its users. The transmitted signal vector from $M$ antennas over one symbol period can be written as
$\textbf{x}=\sum_{j=1}^{M}\sqrt{P_j}\textbf{w}_js_j$,
where $s_j$ denotes the information symbol for the $j$th selected user.
We assume that the transmission power $P_T$ is constant and equally allocated to difference active
beams. Specifically, if $j$ beams are active,
the transmission power allocated to each beam is
$P_j= \frac{P_T}{j}$.

\subsection{Channel Model}

We adopt a log-distance path loss plus Rayleigh block slow fading channel model for the operating environment while ignoring the shadowing effect \cite{gold}. In particular,
the channel gain between the BS and the sensor remains constant over one channel coherence time, denoted by $T_c$, and changes to an independent value afterward.
Let $\textbf{h}_e=[h_{e_1},h_{e_2},\dots,h_{e_M}]^T$ denote the fading channel gain vector from the BS to the sensor,  where $h_{e_m}\in \mathcal{CN}(0, 1)$.
Then the harvested energy at the sensor from the $m$th beam can be given by
\begin{eqnarray}
\label{inter}
E_{m} = \bigg(\frac{\eta T_c}{\Gamma d_H^\lambda}\bigg)\bigg(\frac{P_T}{m}\bigg)|\textbf{h}^T_e\textbf{w}_m|^2, &m=1,2,\dots, M,
\end{eqnarray}
where $d_H$ is the distance from BS to the sensor,
$\eta$ is the energy harvesting efficiency, $\lambda$ is the path loss exponent, ranging from 2 to 5, and $\Gamma$
is a constant parameter of the log-distance model. Specifically, $\Gamma=\frac{PL(d_0)}{d_0^\lambda}
$, where $d_0$ is a reference distance in the antenna far field, and $PL(d_0)$ is linear path loss at distance $d_0$, depending on the propagation environment.
For notational conciseness, we use $\alpha_m$ to denote the amplitude square of the projection of $\textbf{h}_e$ on to
$\textbf{w}_m$, i.e. $\alpha_m=|\textbf{h}^T_e\textbf{w}_m|^2$, whose probability density function (PDF) for Rayleigh fading channel under consideration is given by
\begin{eqnarray}
\label{ray}
f_{\alpha_m}(x)=e^{-x}. 
\end{eqnarray}


\subsection{RUB-based Cooperative Energy Harvesting}

With the proposed cooperative energy harvesting scheme,
the BS will select a maximal number of active beams to serve its users, while ensuring that the harvested energy
at the sensor node during each coherence time is above a predefined energy threshold $E_{th}$.

At the beginning of each channel coherence time,
the BS first estimates the channel vector from the BS to the sensor.
The BS
then calculates and ranks the projection amplitude square $\alpha_m$ for each beam, the order version of which is denoted by $\alpha_{m:M}$, where $\alpha_{1:M}\ge\alpha_{2:M}\ge\dots\ge\alpha_{M:M}$.
After that, the BS calculates the total amount of RF energy that the sensor can harvest when the BS uses the $m$ best beams, corresponding to $\alpha_{1:M}$ to $\alpha_{m:M}$.
The total harvested energy, dented by $E_H$, can be given by
$E_H=\sum_{i=1}^{m}E_{i,m}$,
where $E_{i,m}$ denotes the harvested energy from the $i$th best beam with projection amplitude square $\alpha_{i:M}$, when $m$ best beams are used for transmission, given by
\begin{eqnarray}
\label{rank_e}
E_{i,m}=
\bigg(\frac{\eta T_c}{\Gamma d_H^\lambda}\bigg)\bigg(\frac{P_T}{m}\bigg)\alpha_{i:M}, &i=1,2,\dots,m.
\end{eqnarray}
If the harvested energy with $m$ best beams is larger than the predefined energy threshold $E_{th}$,
whereas the harvested energy with $m+1$ best beams
is less than $E_{th}$, i.e. $\sum_{i=1}^{m}E_{i,m}\ge E_{th}$, and $\sum_{i=1}^{m+1}E_{i,m+1}<E_{th}$,
the BS then uses the $m$ selected beams to serve its users.
Note that with constant total transmission power used at the BS and uniform power allocation, the sensor can harvest more energy from less active beams as the transmission power concentrates on the better beams, i.e. with larger projection power.
It is worth noting that the amount of harvested energy at the sensor may be smaller than $E_{th}$ even when the BS allocates all transmission power $P_T$ to beam $j^*$ corresponding to $\alpha_{1:M}$, i.e. $j^*=\arg\max_j(|\textbf{h}_e^T\textbf{w}_j|^2)$. In this case, the BS will still use beam $j^*$ with transmission power $P_T$ to charge the sensor.

\section{Performance Analysis for the Sensor Node}
\label{scheme}

Intuitively, larger energy threshold $E_{th}$ will lead to a larger amount of harvested energy at the sensor, as well as
a smaller number of served users and possibly smaller sum-rate for the multiuser MIMO system.
In what follows, to evaluate the energy harvesting performance,
we derive the exact statistical distribution of the harvested energy over one coherence time $T_c$ at the sensor, based on which we further calculate the closed form expression of the average harvested energy.

\subsection{Average harvested energy}


The average harvested energy over one coherence time, denoted by $\overline{E}_H$, can be given by
\begin{eqnarray}
\label{exp}
\overline{E}_H&=&
\int_{0}^{\infty}xf_{E_H}(x)dx,
\end{eqnarray}
where $f_{E_H}(x)$ denotes the PDF of harvested energy over one coherence time. In the Appendix A, we derive the close form expression of $f_{E_H}(x)$ as
{\small{
\begin{eqnarray}
\label{pdf}
\nonumber
&&
\hspace{-0.25in}
f_{E_H}(x)
=
\sum_{m=1}^{M-1}
\Bigg\{
\sum_{i=1}^{M-m-1}
\frac{(-1)^{i+1}M!
}{(M-m-1-i)!m!(m-1)!i!}
\sum_{j=0}^{m-1}
\binom{m-1}{j}
(-m)^{m-1-j}
\sum_{t=0}^{j}\frac{j!}{(j-t)!}
\bigg\{
\frac{(m-1-j)!}{(i+1)^{m-j}}\bigg(\frac{m\mu}{E_{th}}\bigg)^{j-t}
\\\nonumber
&&
e^{-\frac{m\mu x}{E_{th}}}x^{j-t-1}\bigg(j-t-\frac{m\mu x}{E_{th}}\bigg)
-
\bigg(\frac{m\mu}{E_{th}}\bigg)^{j-t}
e^{-(i+1)(m+1)\mu}
\sum_{r=0}^{m-j-1}\frac{(m-1-j)!}{(m-1-j-r)!(i+1)^{r+1}}
\sum_{u=0}^{m-1-j-r}(-1)^u
\\\nonumber
&&
\binom{m-1-j-r}{u}
\bigg(\frac{m\mu}{E_{th}}\bigg)^u[(m+1)\mu]^{m-1-j-r-u}
e^{im\mu\frac{x}{E_{th}}}x^{j-t+u-1}
\bigg(j-t+u+\frac{im\mu x}{E_{th}}\bigg)
+e^{-(i+1)(m+1)\mu}
\\\nonumber
&&
\sum_{s=0}^{j-t}
\binom{j-t}{s}
(-1)^s\big[(m+1)\mu\big]^{j-t-s}
\sum_{r=0}^{m-1-j+s}\frac{(m-1-j+s)!}{(m-1-j+s-r)!i^{r+1}}
\sum_{u=0}^{m-1-j+s-r}(-1)^u
\binom{m-1-j+s-r}{u}
\\\nonumber
&&
\bigg(\frac{m\mu}{E_{th}}\bigg)^u[(m+1)\mu]^{m-1-j+s-r-u}
x^{u-1}e^{im\mu\frac{x}{E_{th}}}
\bigg(u+\frac{im\mu}{E_{th}}x\bigg)
\bigg\}
-\frac{M!
}{(M-m-1)!m!(m-1)!}
\sum_{j=0}^{m-1}
\binom{m-1}{j}
(-m)^{m-1-j}
\\\nonumber
&&
\sum_{t=0}^{j}\frac{j!}{(j-t)!}
\bigg\{e^{-(m+1)\mu}
\sum_{s=0}^{j-t}
\binom{j-1}{s}
(-1)^{s}\frac{m\mu}{E_{th}}\big[(m+1)\mu\big]^{j-t-s}\bigg[(m+1)\mu-\frac{m\mu x}{E_{th}}\bigg]^{m-1-j+s}
+
(m-1-j)!\bigg(\frac{m\mu}{E_{th}}\bigg)^{j-t}
\\\nonumber
&&
e^{-m\mu\frac{x}{E_{th}}}x^{j-t-1}\bigg(j-t-\frac{m\mu}{E_{th}}x\bigg)
-
\bigg(\frac{m\mu}{E_{th}}\bigg)^{j-t}
e^{-(m+1)\mu}
\sum_{r=0}^{m-j-1}\frac{(m-1-j)!}{(m-1-j-r)!}
\sum_{u=0}^{m-1-j-r}(-1)^u
\binom{m-1-j-r}{u}
\\\nonumber
&&
\bigg(\frac{m\mu}{E_{th}}\bigg)^u[(m+1)\mu]^{m-1-j-r-u}
(j-t+u)x^{j-t+u-1}
\bigg\}
\Bigg\}
\mathcal{U}\bigg(\Big(1+\frac{1}{m}\Big)E_{th}-x\bigg)\mathcal{U}(x-E_{th})
+\frac{\mu M}{E_{th}}e^{-\frac{\mu x}{E_{th}}}
\\
&&
(1-e^{-\frac{\mu x}{E_{th}}})^{M-1}\mathcal{U}(E_{th}-x)
+
\sum_{s=0}^{M-1}\bigg\{
-(M-1-s)+\frac{M\mu x}{E_{th}}\bigg\}\frac{(M\mu)^{M-1-s}x^{M-2-s}e^{-\frac{M\mu x}{E_{th}}}}{(M-1-s)!E_{th}^{M-1-s}}\mathcal{U}(x-E_{th}),
\end{eqnarray}
}}
where $\mu$ is a constant value equal to $\frac{E_{th}\Gamma d_H^\lambda}{\eta T_1P_T}$ for notational conciseness.
By substituting (\ref{pdf}) into (\ref{exp}) and carrying out integrations, the closed form expression of the average harvested energy over one coherence time can be calculated as
{\footnotesize{
\begin{eqnarray}
\label{exp1}
\nonumber
&&
\hspace{-0.25in}
\overline{E}_H=
\sum_{m=1}^{M-1}
\Bigg\{
-\sum_{i=1}^{M-m-1}
\frac{(-1)^iM!
}{(M-m-1-i)!m!(m-1)!i!}
\sum_{j=0}^{m-1}
\binom{m-1}{j}
(-m)^{m-1-j}
\sum_{t=0}^{j}\frac{j!}{(j-t)!}
\bigg\{
\frac{(m-1-j)!}{(i+1)^{m-j}}\Big(\frac{m\mu}{E_{th}}\Big)^{j-t}
\\\nonumber
&&
\Big\{(j-t)\sum_{r=0}^{j-t}\frac{(j-t)!}{(j-t-r)!}\Big(\frac{E_{th}}{m\mu}\Big)^{r+1}
E_{th}^{j-t-r}e^{-m\mu}
\Big[1-(1+\frac{1}{m})^{j-t-r}e^{-\mu}\Big]
-\frac{m\mu}{E_{th}}
\sum_{r=0}^{j-t+1}\frac{(j-t+1)!}{(j-t+1-r)!}
\\\nonumber
&&
\Big(\frac{E_{th}}{m\mu}\Big)^{r+1}E_{th}^{j-t+1-r}e^{-m\mu}
\Big[1-(1+\frac{1}{m})^{j-t+1-r}e^{-\mu}\Big]\Big\}
-
(\frac{m\mu}{E_{th}})^{j-t}
e^{-(i+1)(m+1)\mu}
\sum_{r=0}^{m-j-1}\frac{(m-1-j)!}{(m-1-j-r)!(i+1)^{r+1}}
\\\nonumber
&&
\sum_{u=0}^{m-1-j-r}(-1)^u
\binom{m-1-j-r}{u}
\Big(\frac{m\mu}{E_{th}}\Big)^u
\big[(m+1)\mu\big]^{m-1-j-r-u}
\Big\{(j-t+u)\sum_{v=0}^{j-t+u}\frac{(j-t+u)!}{(j-t+u-v)!}
(-1)^v\Big(\frac{E_{th}}{im\mu}\Big)^{v+1}
\\\nonumber
&&
E_{th}^{j-t+u-v}e^{im\mu}
\Big[\Big(1+\frac{1}{m}\Big)^{j-t+u-v}e^{i\mu}-1\Big]
+\frac{im\mu}{E_{th}}\sum_{v=0}^{j-t+u+1}\frac{(j-t+u+1)!}{(j-t+u+1-v)!}
(-1)^v\Big(\frac{E_{th}}{im\mu}\Big)^{v+1}E_{th}^{j-t+u+1-v}e^{im\mu}
\\\nonumber
&&
\Big[\Big(1+\frac{1}{m}\Big)^{j-t+u+1-v}e^{i\mu}-1]\Big\}
+e^{-(i+1)(m+1)\mu}\sum_{s=0}^{j-t}
\binom{j-t}{s}
(-1)^s\big[(m+1)\mu\big]^{j-t-s}
\\\nonumber
&&
\sum_{r=0}^{m-1-j+s}\frac{(m-1-j+s)!}{(m-1-j+s-r)!i^{r+1}}
\sum_{u=0}^{m-1-j+s-r}(-1)^u
\binom{m-1-j+s-r}{u}
\Big(\frac{m\mu}{E_{th}}\Big)^u\big[(m+1)\mu\big]^{m-1-j+s-r-u}
\\\nonumber
&&
\Big\{u\sum_{v=0}^{u}\frac{u!}{(u-v)!}(-1)^v\Big(\frac{E_{th}}{im\mu}\Big)^{v+1}E_{th}^{u-v}e^{im\mu}
\Big[\Big(1+\frac{1}{m}\Big)^{u-v}e^{i\mu}-1\Big]
+\frac{im\mu}{E_{th}}\sum_{v=0}^{u+1}\frac{(u+1)!}{(u+1-v)!}(-1)^v
\Big(\frac{E_{th}}{im\mu}\Big)^{v+1}
\\\nonumber
&&
E_{th}^{u+1-v}e^{im\mu}
\Big[\Big(1+\frac{1}{m}\Big)^{u+1-v}e^{i\mu}-1\Big]\Big\}
\bigg\}
-\frac{M!
}{(M-m-1)!m!(m-1)!}
\sum_{j=0}^{m-1}
\binom{m-1}{j}
(-m)^{m-1-j}
\\\nonumber
&&
\sum_{t=0}^{j}\frac{j!}{(j-t)!}
\Big\{e^{-(m+1)\mu}
\sum_{s=0}^{j-t}
\binom{j-t}{s}
(-1)^{s}\frac{m\mu}{E_{th}}\big[(m+1)\mu\big]^{j-t-s}
\sum_{u=0}^{m-1-j+s}(-1)^u
\binom{m-1-j+s}{u}
\Big(\frac{m\mu}{E_{th}}\Big)^{u}
\\\nonumber
&&
[(m+1)\mu]^{m-1-j+s-u}
\frac{E_{th}^{u+2}}{u+2}\Big[\Big(1+\frac{1}{m}\Big)^{u+2}-1\Big]
+(m-1-j)!(\frac{m\mu}{E_{th}})^{j-t}
\Big\{(j-t)
\sum_{v=0}^{j-t}\frac{(j-t)!}{(j-t-v)!}\Big(\frac{E_{th}}{m\mu}\Big)^{v+1}
\\\nonumber
&&
E_{th}^{j-t-v}e^{-m\mu}
\Big[1-\Big(1+\frac{1}{m}\Big)^{j-t-v}e^{-\mu}\Big]
-\frac{m\mu}{E_{th}}\sum_{v=0}^{j-t+1}\frac{(j-t+1)!}{(j-t+1-v)!}\Big(\frac{E_{th}}{m\mu}\Big)^{v+1}E_{th}^{j-t+1-v}e^{-m\mu}
\\\nonumber
&&
\Big[1-\Big(1+\frac{1}{m}\Big)^{j-t+1-v}e^{-\mu}\Big]\Big\}
-
\Big(\frac{m\mu}{E_{th}}\Big)^{j-t}
e^{-(m+1)\mu}
\sum_{r=0}^{m-j-1}\frac{(m-1-j)!}{(m-1-j-r)!}
\sum_{u=0}^{m-1-j-r}(-1)^u
\binom{m-1-j-r}{u}
\\\nonumber
&&
\Big(\frac{m\mu}{E_{th}}\Big)^u[(m+1)\mu]^{m-1-j-r-u}
\frac{j-t+u}{j-t+u+1}
E_{th}^{j-t+u+1}
\Big[\Big(1+\frac{1}{m}\Big)^{j-t+u+1}-1\Big]
\Big\}
\Bigg\}
+
\frac{\mu M}{E_{th}}\sum_{s=0}^{M-1}(-1)^s
\binom{M-1}{s}
\\\nonumber
&&
\frac{E_{th}^2}{(s+1)\mu}\bigg\{\frac{1-e^{-(s+1)\mu}}{(s+1)\mu}-e^{-(s+1)\mu}\bigg\}
+
\sum_{s=0}^{M-1}\frac{(M\mu)^{M-1-s}}{(M-1-s)!E_{th}^{M-1-s}}
\bigg\{
-(M-1-s)\sum_{v=0}^{M-1-s}\frac{(M-1-s)!}{(M-1-s-v)!}
\\
&&
\Big(\frac{E_{th}}{M\mu}\Big)^{v+1}E_{th}^{M-1-s-v}e^{-M\mu}
+\frac{M\mu}{E_{th}}
\sum_{v=0}^{M-s}\frac{(M-s)!}{(M-s-v)!}\Big(\frac{E_{th}}{M\mu}\Big)^{v+1}E_{th}^{M-s-v}e^{-M\mu}
\bigg\}.
\end{eqnarray}
}
}


\subsection{Numerical examples}

We assume the same system parameters for RF energy harvesting as in \cite{mypaper}.
In particular, the transmission power of the BS is $P_T=10kW$. The distance from BS to the sensor
is $d_H=100$ meters.
The pass loss exponent $\lambda$ is assumed to be $3$, and the channel coherence time $T_c$ be $100ms$.

\begin{figure}
\centering
\includegraphics[width=3.5in]{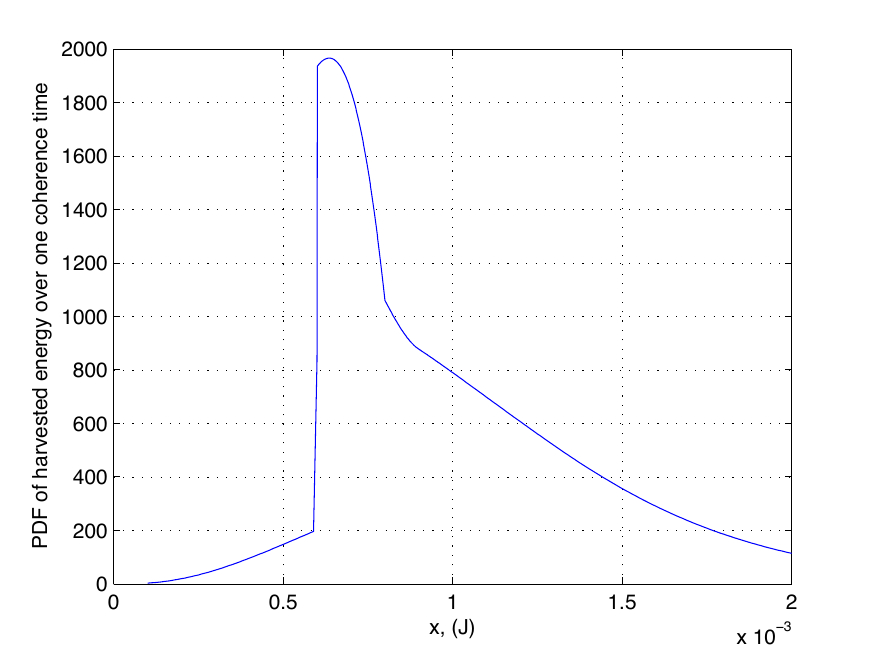}
\caption{Distribution of harvested energy at the sensor.}
\label{pdf_fig}
\end{figure}

In Fig. \ref{pdf_fig}, we plot the PDF of the harvested energy with $E_{th}=0.006J$ for $M=4$ antennas.
As we can see, the harvested energy concentrates around the energy threshold $E_{th}$, as expected.

\begin{figure}
\centering
\includegraphics[width=3.5in]{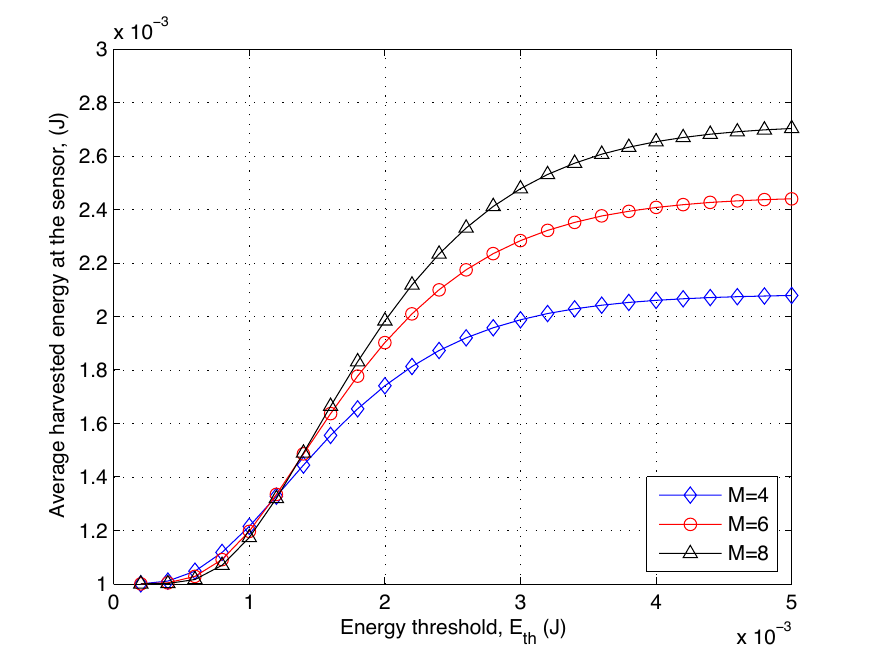}
\caption{Average harvested energy at the sensor.}
\label{exp_fig}
\end{figure}

In Fig. \ref{exp_fig}, we plot the average harvested energy $\overline{E}_h$ as a function of the energy threshold $E_{th}$ for different antenna number $M$.
We can see the average harvested energy at the sensor quickly increases as $E_{th}$ increased, and gradually converges to a constant value when $E_{th}$ is large. This is because when $E_{th}$ is large enough, the BS will only use the best beam to charge the sensor.
We also observe that more antennas leads to smaller average harvested energy when $E_{th}$ is small, and larger average harvested energy when $E_{th}$ is large. This is because when $E_{th}$ is small, more antennas leads to more potential active beams,
which leads to wider distribution of BS transmit power.
When $E_{th}$ is large, the sensor can enjoy more benefits from best beam selection. When $E_{th}$ is 0, the MIMO system serves its users with all beams, and the amount of energy that the sensor can harvest is the same as \cite{mypaper} without considering energy sensitivity and storage capacity.




\section{Performance Analysis for Multiuser MIMO System}
\label{secondary}

In this section, we will analyze the performance of the multiuser MIMO system with proposed cooperative energy harvesting scheme. We first calculate the distribution of the number of active beams that the BS can use for transmission, based on which we further carry out accurate sum-rate analysis for the multiuser MIMO system.

\subsection{Distribution of the number of active beams}
In the following, we derive the probability mass function (PMF) of the number of active beams that the BS can use for transmission, $M_a$, which will be applied to the sum-rate analysis of the existing system.
According to the RUB-based cooperative energy harvesting scheme, the PMF of the number of active beams $M_a$
can be mathematically given by
\begin{eqnarray}
\label{act_dis}
\Pr[M_a=m] &=&
\begin{cases}
\Pr\bigg[\sum_{i=1}^{m}E_{i,m}\ge E_{th},
\sum_{i=1}^{m+1}E_{i,m+1}<E_{th}\bigg],
&1< m< M,
\\
\Pr\bigg[\sum_{i=1}^{M}E_{i,M}\ge E_{th}\bigg],&m=M,
\end{cases}
\end{eqnarray}
which can be further calculated as
\begin{eqnarray}
\label{act_dis}
\Pr[M_a=m] &=&
\begin{cases}
\int_{0}^{\mu}\int_{m\mu}^{(m+1)\mu-x}f_{\alpha_{m+1:M},z_m}(x,y)dxdy
,
&1< m< M,
\\
\int_{m\mu}^{\infty}f_{z_m}(x)dx,
&m=M,
\end{cases}
\end{eqnarray}
where
$f_{z_m}(x)$, and $f_{\alpha_{m+1}, z_m}(x)$ are the marginal and joint PDFs of $\alpha_{m+1:M}$ and $z_m$.
Specifically, the BS can use only one beam for transmission when the harvested RF energy at the sensor from the best beam $j^*$, i.e. $E_{1,1}$
is less than $E_{th}$, or when the harvested RF energy $E_{1,1}$ is larger than $E_{th}$, and $\sum_{i=1}^{2}E_{i,2}$ is less than $E_{th}$. Then the probability that only one beam is active can be mathematically given by
\begin{eqnarray}
\label{act_dis1}
\Pr[M_a=1] &=&
\Pr[E_{1,1}< E_{th}]+
\Pr[E_{1,1}\ge E_{th}, E_{1,2}+E_{2,2}< E_{th}],
\end{eqnarray}
which can be calculated as
\begin{eqnarray}
\label{act_dis1}
\Pr[M_a=1]
&=&
\int_{0}^{\mu}f_{z_1}(x)dx+\int_{0}^{\mu}\int_{\mu}^{2\mu-x}f_{\alpha_{2:M},z_1}(x,y)dxdy.
\end{eqnarray}
By substituting (\ref{large}) into (\ref{act_dis1}), (\ref{small}) and (\ref{nei}) into (\ref{act_dis}), and carrying out integrations, the closed form of the probability mass function (PMF) of the active beam number can be calculated as 
\begin{eqnarray}
\label{1beam_number}
\Pr[M_a=m] =
\begin{cases}
(1-e^{-\mu})^M
+
\sum_{i=1}^{M-2}
\frac{(-1)^iM!
}{(M-2-i)!i!}
\bigg\{
\frac{e^{-\mu}}{i+1}\Big[1-e^{-(i+1)\mu}
\Big]
\\
-\frac{e^{-2\mu}}{i}
\Big[1-e^{-i\mu}
\Big]
\bigg\}
+\frac{M!
}{(M-2)!}
\bigg\{
e^{-\mu}
(1
-e^{-\mu}
)
-\mu e^{-2\mu}
\bigg\}
, &m=1,\\
\sum_{i=1}^{M-m-1}
\frac{(-1)^iM!
}{(M-m-1-i)!m!(m-1)!i!}
\sum_{j=0}^{m-1}
\binom{m-1}{j}
(-m)^{m-1-j}
\\
\sum_{t=0}^{j}\frac{j!}{(j-t)!}
\bigg\{
e^{-m\mu}(m\mu)^{j-t}
\Big[\frac{(m-1-j)!}{(i+1)^{m-j}}
\\
-\sum_{r=0}^{m-j-1}\frac{(m-1-j)!\mu^{m-1-j-r}e^{-(i+1)\mu}}{(m-1-j-r)!(i+1)^{r+1}}
\Big]
\\
-e^{-(m+1)\mu}\sum_{s=0}^{j-t}
\binom{j-t}{s}
(-1)^s\big[(m+1)\mu\big]^{j-t-s}
\\
\Big[\frac{(m-1-j+s)!}{i^{m-j+s}}-\sum_{r=0}^{m-1-j+s}\frac{(m-1-j+s)!\mu^{m-1-j+s-r}e^{-i\mu}}{(m-1-j+s-r)!i^{r+1}}
\Big]
\bigg\}
\\
+\frac{M!
}{(M-m-1)!m!(m-1)!}
\sum_{j=0}^{m-1}
\binom{m-1}{j}
(-m)^{m-1-j}
\\
\sum_{t=0}^{j}\frac{j!}{(j-t)!}
\bigg\{
e^{-m\mu}(m\mu)^{j-t}
\Big[(m-1-j)!
\\
-\sum_{r=0}^{m-j-1}\frac{(m-1-j)!\mu^{m-1-j-r}e^{-\mu}}{(m-1-j-r)!}
\Big]
\\
-e^{-(m+1)\mu}\sum_{s=0}^{j-t}
\binom{j-t}{s}
(-1)^s\big[(m+1)\mu\big]^{j-t-s}
\frac{\mu^{m-j+s}}{m-j+s}
\bigg\}
,
&1< m< M,
\\
\sum_{s=0}^{m-1}\frac{e^{-m\mu}(m\mu)^{m-1-s}}{(m-1-s)!}
,&m=M.
\end{cases}
\end{eqnarray}

\begin{figure}
\centering
\includegraphics[width=3.5in]{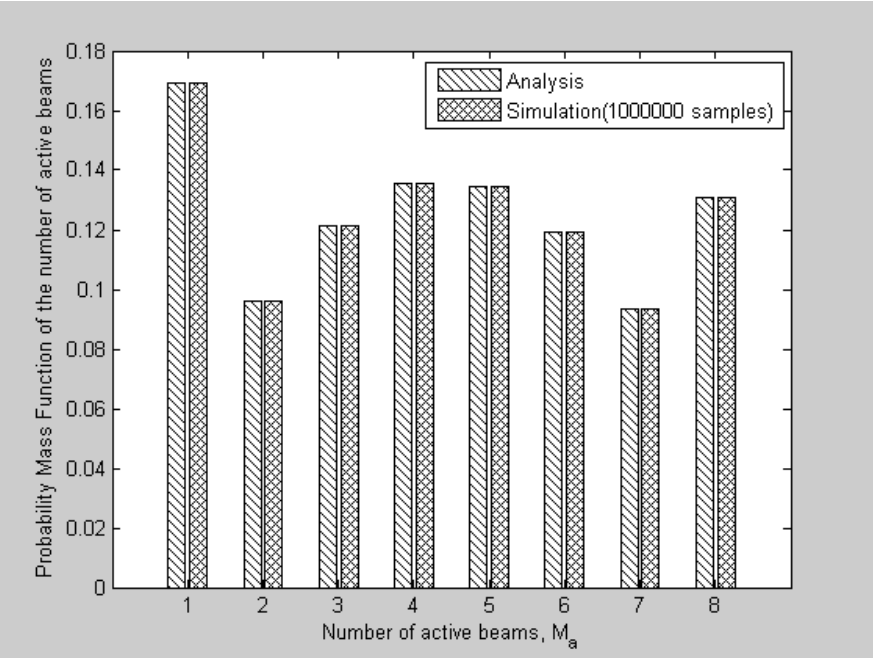}
\caption{Distribution of the number of active beams.}
\label{pmf}
\end{figure}

In Fig. \ref{pmf}, we plot the probability mass function (PMF) of the number of active beams needed for $M=8$ antennas. The PMF obtained through monte carlo simulation are also presented. As we can see, the analytical result matches the simulation result perfectly, which verify our analytical derivation.

\subsection{Sum-rate of multiuser MIMO system}

The average sum-rate of the multiuser MIMO system can be calculated as
\begin{equation}
\label{rate}
R=\sum_{i=1}^M\Pr[M_a=i]R_{M_a},
\end{equation}
where $\Pr[M_a=i]$ represents the probability that $M_a$ beams are active, which has been given in (\ref{1beam_number}), $R_{M_a}$ denotes the average sum-rate of the existing system conditioning on $M_a$ beams are active, which can be calculated using the user selection scheme in \cite{hcy1}.
\begin{figure}
\centering
\includegraphics[width=3.5in]{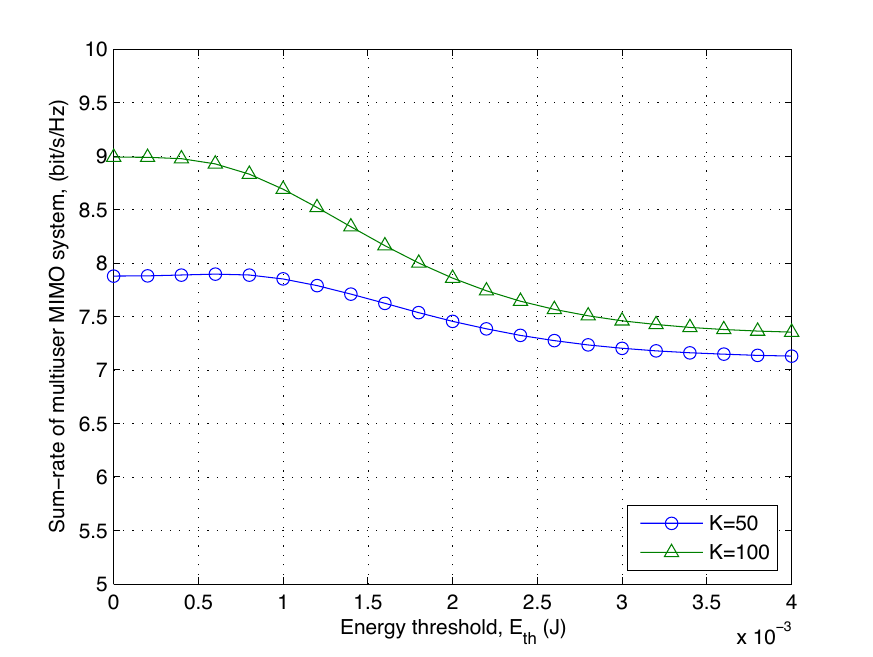}
\caption{Sum-rate of multiuser MIMO system for M=4 antennas.}
\label{rate_rub}
\end{figure}



In Fig. \ref{rate_rub}, we plot the average sum-rate of the multiuser MIMO system as a function of the energy threshold $E_{th}$ for different user number $K$ for $M$=4 antennas, assuming the user selection scheme proposed in \cite{hcy1}.
We can observe that larger user number leads to larger sum-rate, due to user selection.
We also observe that the sum-rate reduces gradually to a constant value with the increase of $E_{th}$. This is because when $E_{th}$ is large, the BS will only use the best beam, from the energy harvesting perspective, to serve its selected user.
Combined with Fig. \ref{exp_fig}, we can see there exists a tradeoff
of average harvested energy at the sensor versus sum-rate
of the multiuser MIMO system.
In particular, larger $E_{th}$ leads to larger average harvested energy, but smaller sum-rate. We can achieve desired energy harvesting performance  by properly adjusting $E_{th}$ at the expense of certain sum-rate degradation.

\section{Conclusion}
\label{conclude}
We proposed a RUB-based cooperative beam selection scheme, using which
the existing multiuser MIMO system can help increase the amount of harvested energy of wireless sensor nodes.
We obtained the closed form expression of the distribution of harvested energy and the average harvested energy of the sensor node, based on which, we investigate the tradeoff of the average harvested energy versus the sum-rate of the multiuser MIMO system.
The analytical results will greatly help improve the performance of
sensor implementation powered by RF energy harvesting for the appropriate sensing applications.

\appendices
\section{Distribution of Harvested Energy over One Coherence Time}

Conditioning on the number of active beams for transmission, denoted by $M_a$, the cumulative distribution function (CDF) of $E_H$ can be represented as
\begin{eqnarray}
\label{cdf}
F_{E_H}(x)
=
\sum_{m=1}^{M}\Pr[E_H<x, M_a=m].
\end{eqnarray}
According to our proposed cooperative beam selection scheme, the number of active beams $M_a$
is equal to $m$ $(1<m<M)$ if and only if $\sum_{i=1}^{m}E_{i,m}\ge E_{th}$, and $\sum_{i=1}^{m+1}E_{i,m+1}<E_{th}$.
Furthermore, the number of active beams $M_a$ is equal to 1 if the energy threshold $E_{th}$ can not be satisfied with all transmission power $P_T$ allocated to the best beam, i.e., $E_{1,1}<E_{th}$, and equal to $M$ if
the harvested energy is larger than $E_{th}$ with $P_T$ allocated to all $M$ beams, i.e.,
$\sum_{i=1}^{M}E_{i,M}\ge E_{th}$.
Therefore, we can rewrite (\ref{cdf}) as
\begin{eqnarray}
\label{cdf1}
\nonumber
F_{E_H}(x)
&=&
\sum_{m=1}^{M-1}
\Pr\bigg[\sum_{i=1}^{m}E_{i,m}<x, \sum_{i=1}^{m}E_{i,m}\ge E_{th},
\sum_{i=1}^{m+1}E_{i,m+1}<E_{th}\bigg]
\\
&&
+\Pr[E_{1,1}<x, E_{1,1}<E_{th}]
+\Pr\bigg[\sum_{i=1}^{M}E_{i,M}<x, \sum_{i=1}^{M}E_{i,M}\ge E_{th}\bigg].
\end{eqnarray}
For notational conciseness,
we denote the sum of the $m$ largest projection amplitude square from $\alpha_{1:M}$ to $\alpha_{m:M}$ as $z_m$, i.e. $z_m=\sum_{i=1}^{m}\alpha_{i:M}$. It follows that $\sum_{i=1}^{m}E_{i,m}=
\bigg(\frac{\eta T_c}{\Gamma d_H^\lambda}\bigg)\bigg(\frac{P_T}{m}\bigg)z_m$.
Then (\ref{cdf1}) can be mathematically calculated as
\begin{eqnarray}
\label{cdf2}
\nonumber
F_{E_H}(x)
&=&
\sum_{m=1}^{M-1}\Bigg\{
\int_{0}^{\mu}\int_{m\mu}^{(m+1)\mu-y}f_{\alpha_{m+1:M},z_m}(y,z)dydz
-\int_{0}^{(m+1)\mu-\frac{m\mu x}{E_{th}}}\int_{\frac{m\mu x}{E_{th}}}^{(m+1)\mu-y}
\\
&&
f_{\alpha_{m+1:M},z_m}(y,z)dydz\Bigg\}
+\int_{0}^{\frac{\mu x}{E_{th}}}f_{z_1}(y)dy
+\int_{M\mu}^{\frac{M\mu x}{E_{th}}}f_{z_M}(y)dy,
\end{eqnarray}
where  $\mu$ is a constant value equal to $\frac{E_{th}\Gamma d_H^\lambda}{\eta T_1P_T}$ for notational conciseness, $f_{z_1}(x)$, $f_{z_m}(x)$, and $f_{\alpha_{m+1}, z_m}(x)$ denote the PDF of $z_1$, $z_m$, and the joint PDF of $\alpha_{m+1}$ and $z_m$, respectively, the closed-form expression of which can be obtained as \cite{hcy},
\begin{eqnarray}
\label{large}
f_{z_1}(x) = Me^{-x}(1-e^{-x})^{M-1},
\end{eqnarray}
\begin{eqnarray}
\label{small}
\nonumber
f_{z_m}(x)&=&\frac{M!}{(M-m)!m!}e^{-x}\Bigg[\frac{x^{m-1}}{(m-1)!}
+\sum_{j=1}^{M-m}(-1)^{m+j-1}
\frac{(M-m)!}{(M-m-j)!j!}
\\
&&
\Big(\frac{m}{j}\Big)^{m-1}
\bigg(e^{-\frac{jx}{m}}-\sum_{t=0}^{m-2}\frac{1}{t!}\Big(-\frac{jx}{m}\Big)^t
\bigg)
\Bigg]
,
\end{eqnarray}
and
\begin{eqnarray}
\label{nei}
f_{\alpha_{m+1:M},z_m}(x, y) &=&
\sum_{i=0}^{M-m-1}
\frac{(-1)^iM!(y-mx)^{m-1}e^{-y-(i+1)x}
}{(M-m-1-i)!m!(m-1)!i!}
, \ \  y\ge mx,
\end{eqnarray}
respectively.
By substituting (\ref{large})(\ref{small}) and (\ref{nei}) into (\ref{cdf}) and carrying out integrations, the CDF of the harvested energy over one coherence time can be calculated as 
{\footnotesize{
\begin{eqnarray}
\label{cdf_close}
\nonumber
&& 
\hspace{-0.3in}
F_{E_H}(x)
=
(1-e^{-\mu\frac{x}{E_{th}}})^M
\mathcal{U}(E_{th}-x)
+
\Bigg\{1-\sum_{s=0}^{M-1}\frac{e^{-M\mu}(M\mu)^{M-1-s}}{(M-1-s)!}-(1-e^{-\mu})^M
-
\sum_{m=1}^{M-1}
\sum_{i=1}^{M-m-1}
\\\nonumber
&&
\frac{(-1)^iM!
}{(M-m-1-i)!m!(m-1)!i!}
\sum_{j=0}^{m-1}
\binom{m-1}{j}
(-m)^{m-1-j}
\sum_{t=0}^{j}\frac{j!}{(j-t)!}
\bigg\{
e^{-m\mu\frac{x}{E_{th}}}(m\mu\frac{x}{E_{th}})^{j-t}
\Big[\frac{(m-1-j)!}{(i+1)^{m-j}}
\\\nonumber
&&
-\sum_{r=0}^{m-j-1}\frac{(m-1-j)![(m+1)\mu-m\mu\frac{x}{E_{th}}]^{m-1-j-r}e^{-(i+1)[(m+1)\mu-m\mu\frac{x}{E_{th}}]}}{(m-1-j-r)!(i+1)^{r+1}}
\Big]
-e^{-(m+1)\mu}\sum_{s=0}^{j-t}
\binom{j-t}{s}
\\\nonumber
&&
(-1)^s\big[(m+1)\mu\big]^{j-t-s}
\Big[\frac{(m-1-j+s)!}{i^{m-j+s}}
-\sum_{r=0}^{m-1-j+s}\frac{(m-1-j+s)!}{(m-1-j+s-r)!i^{r+1}}
e^{-i[(m+1)\mu-m\mu\frac{x}{E_{th}}]}
\\\nonumber
&&
[(m+1)\mu-m\mu\frac{x}{E_{th}}]^{m-1-j+s-r}
\Big]
\bigg\}
+\frac{M!
}{(M-m-1)!m!(m-1)!}
\sum_{j=0}^{m-1}
\binom{m-1}{j}
(-m)^{m-1-j}
\sum_{t=0}^{j}\frac{j!}{(j-t)!}
\\\nonumber
&&
\bigg\{
e^{-m\mu\frac{x}{E_{th}}}\Big(m\mu\frac{x}{E_{th}}\Big)^{j-t}
\Big[(m-1-j)!
-\sum_{r=0}^{m-j-1}\frac{(m-1-j)!}{(m-1-j-r)!}
\Big[(m+1)\mu-m\mu\frac{x}{E_{th}}\Big]^{m-1-j-r}
\\\nonumber
&&
e^{-[(m+1)\mu-m\mu\frac{x}{E_{th}}]}
\Big]
-e^{-(m+1)\mu}
\sum_{s=0}^{j-t}
\binom{j-t}{s}
(-1)^{s}\big[(m+1)\mu\big]^{j-t-s}
\frac{[(m+1)\mu-m\mu\frac{x}{E_{th}}]^{m-j+s}}{m-j+s}
\bigg\}
\Bigg\}
\\\nonumber
&&
\mathcal{U}\bigg(\Big(1+\frac{1}{m}\Big)E_{th}-x\bigg)
\mathcal{U}(x-E_{th})
\Bigg\}
+\Bigg\{(1-e^{-\mu})^M
+
\sum_{s=0}^{M-1}\bigg\{e^{-M\mu}
-e^{-M\mu\frac{x}{E_{th}}}\Big(\frac{x}{E_{th}}\Big)^{M-1-s}\bigg\}
\\
&&
\frac{(M\mu)^{M-1-s}}{(M-1-s)!}
\Bigg\}
\mathcal{U}(x-E_{th}).
\end{eqnarray}
}}
After taking derivative with respect to x, the PDF of $F_{E_H}(x)$ is derived and given in (\ref{pdf}).

\end{document}